\documentclass[aps,prl,reprint,superscriptaddress,longbibliography]{revtex4-2}

\usepackage[T1]{fontenc}
\usepackage{lmodern}
\usepackage{microtype}
\usepackage{amsmath,amssymb,mathtools}
\usepackage{booktabs}
\usepackage{graphicx}
\usepackage[dvipsnames]{xcolor}
\usepackage{pgfplots}
\pgfplotsset{compat=1.14}
\usepackage[colorlinks=true,allcolors=MidnightBlue]{hyperref}
\usepackage{braket}

\newcommand{\dd}{\,\mathrm d}
\newcommand{\Tr}{\operatorname{Tr}}
\newcommand{\ketbra}[2]{\lvert #1\rangle\!\langle #2\rvert}
\newcommand{\R}{\mathbb R}

\begin{document}
	
\title{Wigner-positive quantum states can have lower entropy than the vacuum}

\author{Zacharie Van Herstraeten}
\email{zacharie.van-herstraeten@inria.fr}
\affiliation{DIENS, École Normale Supérieure, PSL  University, CNRS, INRIA (QAT), 45 rue d’Ulm, Paris, 75005, France}

\author{Nicolas J. Cerf}
\affiliation{Centre for Quantum Information and Communication, École polytechnique de Bruxelles, CP 165, Université libre de Bruxelles, 1050 Brussels, Belgium}

\author{Ulysse Chabaud}
\email{ulysse.chabaud@inria.fr}
\affiliation{DIENS, École Normale Supérieure, PSL  University, CNRS, INRIA (QAT), 45 rue d’Ulm, Paris, 75005, France}

\begin{abstract}
The Wigner entropy conjecture posits that pure Gaussian states minimize the Shannon entropy of non-negative Wigner functions, known as the Wigner entropy. We show that mixing a specific pure Wigner-negative state with another suitably chosen quantum state can restore Wigner positivity while retaining a Wigner entropy that is slightly -- but strictly -- below the vacuum entropy.
As a consequence, we disprove the Wigner entropy conjecture, as well as the stronger Wigner majorization conjecture, by deriving simple analytical counterexamples with violations of the entropy bound no larger than $10^{-3}$. Such states with sub-vacuum Wigner entropy can be found arbitrarily close to the vacuum and can also exhibit sub-vacuum Wigner--Rényi $\alpha$-entropy for $0<\alpha<2$. We identify the physical mechanism behind the existence of such states, which arises from a subtle interplay between the uncertainty principle and non-Gaussianity. The key idea of this paper was developed with the help of AI tools, building on the characterization of extreme non-negative Wigner functions. 
\end{abstract}

\maketitle

\emph{Introduction.---}Quantum states can be represented via quasi-probability distributions supported on the phase space of position- and momentum-like coordinates $(x,p)$. Quasi-probability distributions integrate to one but, unlike classical probability distributions, they can take negative values.
Among them, the Wigner function is arguably the most common phase-space representation of a quantum state~\cite{wigner1932,hillery1984}. It is defined for a single-mode quantum state $\rho$ in the position basis $\{|x\rangle\}$ as
\begin{equation}
    W_\rho(x,p)=\frac1\pi\int_{y\in\R}e^{-2ipy}\bra{x+y}\rho\ket{x-y}\dd y,
\end{equation}
with the convention $\hbar=1$. Note that there is a one-to-one correspondence between $\rho$ and $W_\rho$. The Wigner function is
central to quantum optics and optical state tomography~\cite{lvovsky2009}, since its marginals yield the position and momentum probability densities $P_x$ and $P_p$, namely
\begin{equation}
    P_x(x)=\int_{p\in\R}W_\rho(x,p)\dd p\,,\quad P_p(p)=\int_{x\in\R}W_\rho(x,p)\dd x.
\end{equation}
The Wigner function is also instrumental in bosonic quantum information theory~\cite{weedbrook2012}, since its negative values may be thought of as a necessary resource for quantum computational advantage~\cite{albarelli2018,booth2022contextuality}, in the sense that quantum computations described by non-negative Wigner functions can be simulated efficiently on a classical computer~\cite{mari2012}. 

When the Wigner function is non-negative, it provides a valid joint position-momentum probability distribution, which nevertheless remains non-trivially constrained by quantum mechanics, that is, the fact that $\rho$ is positive semidefinite. The most notable constraint is the Heisenberg uncertainty relation, namely
\begin{equation}
\label{eq:heisenberg_uncertainty}
\sigma_x^2\cdot\sigma_p^2\ge\frac14,
\end{equation}
where $\sigma_x^2$ and $\sigma_p^2$ are the variances of the marginals $P_x$ and $P_p$ of the Wigner function, implying that the latter cannot be localized beyond a certain limit. The uncertainty relation \eqref{eq:heisenberg_uncertainty} may be strengthened to an entropic uncertainty relation~\cite{beckner1975inequalities,bialynicki1975uncertainty}, namely
\begin{equation}
\label{eq:entropic_uncertainty}
 h(P_x)+h(P_p)\ge1+\ln \pi,
\end{equation}
where $h(P)=-\int P\ln P$ is the (differential) Shannon entropy. Note that for Gaussian states, Eq.~\eqref{eq:entropic_uncertainty} reduces to Eq.~\eqref{eq:heisenberg_uncertainty} because the entropy of a Gaussian probability distribution of variance $\sigma^2$ is equal to $\frac12 \ln(2\pi e \sigma^2)$. 
Hudson's theorem implies that the only pure states with non-negative Wigner function are Gaussian states~\cite{hudson1974,soto1983wigner}, which saturate both Eqs.~\eqref{eq:heisenberg_uncertainty} and \eqref{eq:entropic_uncertainty} whenever $x$ and $p$ are uncorrelated, while the set of mixed states with non-negative Wigner function---called \textit{Wigner-positive} states hereafter---is significantly more involved~\cite{broecker1995,mandilara2009extending,sets2025,extreme2025}.
%Hudson's theorem implies that the only pure states with non-negative Wigner function are Gaussian states~\cite{hudson1974,soto1983wigner}, which then saturate both uncertainty principles, Eqs.~\eqref{eq:heisenberg_uncertainty} and \eqref{eq:entropic_uncertainty}, while the set of mixed states with non-negative Wigner function---called \textit{Wigner-positive} states hereafter---is significantly more involved~\cite{broecker1995,mandilara2009extending,sets2025,extreme2025}.

Since the Wigner function of a Wigner-positive state is a valid joint $(x,p)$ probability distribution, it has a well-defined Shannon entropy, the so-called \emph{Wigner entropy}, defined as~\cite{vanherstraeten2021}
\begin{equation}
 h(W_\rho)=-\int_{\R^2}W_\rho(x,p)\ln W_\rho(x,p)\dd x\dd p,
 \label{eq:entropy}
\end{equation}
for a single-mode Wigner-positive state $\rho$.
The Wigner function of the vacuum state $\ket{0}$ is given by $W_{\ket0\!\bra0}(x,p)=\frac1\pi e^{-(x^2+p^2)}$, with the corresponding Wigner entropy 
\begin{equation}
    h(W_{\ket0\!\bra0})=1+\ln\pi.
\end{equation}
Moreover, since Gaussian unitaries (i.e., displacements, rotations, and squeezing) preserve the Wigner entropy, all pure Gaussian states have a Wigner entropy equal to that of the vacuum. As first explored in Ref.~\cite{hertz2017},
this suggests the possible existence of a phase-space Wigner uncertainty principle analogous to the Wehrl--Lieb inequality for the Husimi function~\cite{wehrl1978,lieb1978}, namely
\begin{equation}
 h(W_\rho) \ge 1 + \ln \pi \, .
\label{Wigner-uncertainty-rel}
\end{equation}
Compared with the entropic uncertainty relation~\eqref{eq:entropic_uncertainty}, this inequality---if proven---would yield a stronger bound whenever $x$ and $p$ are correlated, since 
$h(W_\rho)=h(P_x)+h(P_p)-I(x{:}p)$, where $I(x{:}p)\ge 0$ denotes the mutual information. The trade-off, however, is that, unlike \eqref{eq:entropic_uncertainty}, this inequality applies only to Wigner-positive states. Restricting to such states, the core question can thus be formulated as a fundamental conjecture:

\medskip

\emph{Wigner entropy conjecture~\cite{vanherstraeten2021}.---}Pure Gaussian states minimize the Wigner entropy, that is,
\begin{equation}
 h(W_\rho)\geq 1+\ln\pi,
 \label{eq:wec}
\end{equation}
for every single-mode Wigner-positive state $\rho$.

\medskip

Since the entropy is a concave functional, Conjecture~(\ref{eq:wec}) is trivially verified for any mixture of Gaussian states. In particular, it holds for all mixtures of coherent states (i.e., Glauber's classical states). 
The formulation of this conjecture has been extended to the Rényi entropies of the Wigner function (so-called Wigner--Rényi entropies), 
\begin{equation}
 h_\alpha(W_\rho)=\frac1{1-\alpha}\ln\int_{\R^2}[W_\rho(x,p)]^\alpha\dd x\dd p,
 \label{eq:WRentropy}
\end{equation}
with the limit case $\alpha\to 1$ reducing to the Wigner entropy~\cite{vanherstraeten2021}. The conjecture has been further generalized to the stronger \emph{Wigner majorization conjecture}~\cite{majorization2023}, which asserts that the Wigner function of the vacuum majorizes that of any other Wigner-positive state.

The partial validity of Conjecture~(\ref{eq:wec}) has been established for various families of Wigner-positive states. It was first proven for passive states~\cite{vanherstraeten2021} (i.e., mixtures of Fock states with decreasing weights). It was then shown that the Wigner function of the vacuum majorizes that of every Wigner-positive mixture of the vacuum with the two lowest Fock states $\ket{1}$ and $\ket{2}$, which implies that the conjecture holds for such states~\cite{majorization2023}. This result was subsequently strengthened by proving a similar majorization relation for arbitrary Wigner-positive Fock mixtures, implying the validity of the conjecture for this entire family of states~\cite{vanbever2021}. Other established cases include all Wigner-positive states in the vacuum and one-photon sector $\mathrm{span}\{\ket0,\ket1\}$~\cite{qian2024} as well as all beam-splitter states~\cite{interference2025} (i.e., states obtained by applying a balanced beam-splitter to a separable two-mode input and discarding one output mode). Furthermore, it was shown that  $h(W_\rho)\ge \ln(2\pi) - \ln(\mathrm{tr} (\rho^2))$ \cite{note_ulysse_2021}, hence the conjecture is true for all states that are sufficiently mixed, namely with purity below $\frac2e\approx0.736$~\cite{purity2026}. Finally, the extended conjecture for Wigner--Rényi $\alpha$-entropies was also proven true for $\alpha\ge2$~\cite{diasprata2023}. Despite these results, a full proof of the Wigner entropy conjecture has remained elusive. 

\newcommand{\weccell}[2]{%
  \parbox[t]{#1}{\raggedright\strut #2\strut}%
}

\begin{table*}[t]
  \footnotesize
  \renewcommand{\arraystretch}{1.4}
  \begin{tabular}{@{}lll@{}}
    \toprule
    \weccell{0.38\textwidth}{Family of Wigner-positive states}
    &
    \weccell{0.37\textwidth}{Status of the Wigner entropy conjecture}
    &
    \weccell{0.18\textwidth}{Reference}
    \\
    \midrule
    \weccell{0.38\textwidth}{Pure states}
    &
    \weccell{0.37\textwidth}{Holds (with equality)}
    &
    \weccell{0.18\textwidth}{Ref.~\cite{hudson1974}}
    \\
    \weccell{0.38\textwidth}{Mixed states with purity at most $2/e$}
    &
    \weccell{0.37\textwidth}{Holds}
    &
    \weccell{0.18\textwidth}{Ref.~\cite{purity2026}}
    \\
    \weccell{0.38\textwidth}{Mixtures of Fock states}
    &
    \weccell{0.37\textwidth}{Holds (Wigner majorization holds)}
    &
    \weccell{0.18\textwidth}{Ref.~\cite{vanbever2021}}
    \\
    \weccell{0.38\textwidth}{\quad Passive states}
    &
    \weccell{0.37\textwidth}{Holds}
    &
    \weccell{0.18\textwidth}{Ref.~\cite{vanherstraeten2021}}
    \\
    \weccell{0.38\textwidth}{\quad Mixtures of Fock states in $\operatorname{span}\{\ket0,\ket1,\ket2\}$}
    &
    \weccell{0.37\textwidth}{Holds (Wigner majorization holds)}
    &
    \weccell{0.18\textwidth}{Ref.~\cite{majorization2023}}
    \\
    \weccell{0.38\textwidth}{Beam-splitter states}
    &
    \weccell{0.37\textwidth}{Holds}
    &
    \weccell{0.18\textwidth}{Ref.~\cite{interference2025}}
    \\    
    \weccell{0.38\textwidth}{\quad States in $\mathrm{span}\{\ket0,\ket1\}$}
    &
    \weccell{0.37\textwidth}{Holds}
    &
    \weccell{0.18\textwidth}{Ref.~\cite{qian2024}}
    \\
    \weccell{0.38\textwidth}{\quad States in $\mathrm{span}\{\ket0,\ket2\}$}
    &
    \weccell{0.37\textwidth}{Holds}
    &
    \weccell{0.18\textwidth}{This work}
    \\
    \weccell{0.38\textwidth}{States in $\mathrm{span}\{\ket0,\ket n\}$, $n\geq3$}
    &
    \weccell{0.37\textwidth}{False}
    &
    \weccell{0.18\textwidth}{This work}
    \\
    \weccell{0.38\textwidth}{States in $\mathrm{span}\{\ket0,\ket1,\ket2\}$}
    &
    \weccell{0.37\textwidth}{False}
    &
    \weccell{0.18\textwidth}{This work}
    \\
    \bottomrule
  \end{tabular}
  \caption{Status of the Wigner entropy conjecture \eqref{eq:wec} for
 families (and sub-families) of Wigner-positive states.
  ``Holds'' means that the conjecture is proven for every state in the
  family; ``false'' means that the family contains counterexamples. The inclusion of Wigner-positive states in $\mathrm{span}\{\ket0,\ket1\}$ and $\mathrm{span}\{\ket0,\ket2\}$ within the set of beam-splitter states is proven in this work.}
  \label{tab:wec-status}
\end{table*}

\medskip

\emph{Results.---}In this work, we provide a definitive answer to this problem by showing that the Wigner entropy conjecture does not hold for all Wigner-positive states, i.e., \emph{there exist Wigner-positive states with lower Wigner entropy than the vacuum}. As a direct consequence, the stronger Wigner majorization conjecture is also disproved, and we shall show that even the Wigner--Rényi entropy version of the conjecture is ruled out for $0<\alpha<2$. 

Since the Wigner entropy is concave, searching for counterexamples naturally leads us to consider the boundary of the convex set of Wigner-positive states, and in particular its extreme points \cite{sets2025,extreme2025}. Intuitively, we look for instances in which $h(W_\rho)$ decreases as the state $\rho$ approaches this boundary, possibly falling below $1+\ln\pi$ just before the boundary is reached. More precisely, to construct counterexamples, we exploit the fact that the Wigner entropy can be extended to Wigner-negative states by introducing a complex-valued entropy function ~\cite{complex2024}, whose real part $h_r(W)=-\int W\ln |W|$ is known to fall below $1+\ln\pi$ for certain pure Wigner-negative states. We choose such a Wigner-negative state with only a small Wigner-negative volume, measured by the imaginary part of the complex-valued Wigner entropy $h_i(W)=\pi \int_{W<0} |W|$, so that its negative regions can easily be eliminated. This is done by mixing this state with another suitably chosen quantum state that removes the negative regions, while preserving a sub-vacuum value of the Wigner entropy. 

In what follows, we present two analytical families of Wigner-positive states for which we disprove the conjecture. The first family lies in $\mathrm{span}\{\ket0,\ket n\}$ for any $n\ge3$, while the second family belongs to $\mathrm{span}\{\ket0,\ket 1,\ket 2\}$. We show that these counterexamples are minimal (in terms of Fock support) by proving that the conjecture holds in $\mathrm{span}\{\ket0,\ket 2\}$. In both families, our counterexamples are states that are close to the vacuum, highly pure, and admit a non-zero Fock-coherence, which is consistent with the special cases for which the conjecture holds. Table~\ref{tab:wec-status} summarizes the status of the Wigner entropy conjecture. We refer the reader to Appendix \ref{app:prelim} for preliminary material (conventions, notations, and tools used  in the following). 

\medskip

\begin{figure*}[t]
\centering
\includegraphics[width=0.95\linewidth]{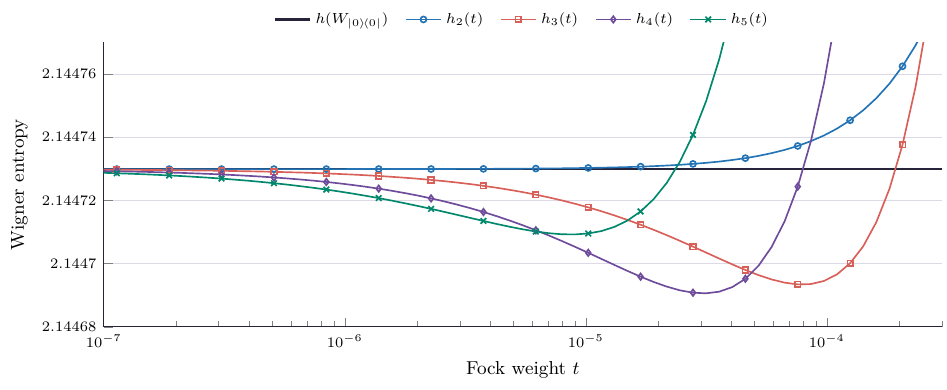}
\caption{\label{fig:results}Wigner entropy of the state $\rho_n(t,s)$ as a function of the mixing parameter $t$, choosing the maximal Fock-coherence $s_n(t)$ such that $\rho_n(t,s)$ is Wigner-positive. The horizontal black line is the Wigner entropy of the vacuum $h(W_{\ket0\!\bra0})=1+\ln\pi$. The $n=2$ curve approaches the vacuum value from above, whereas the $n\ge3$ curves undershoot it for $t$ small enough. The largest violation  ($\simeq 4 \cdot 10^{-5}$) is observed for $n=4$.}
\end{figure*}

\emph{Two-level counterexamples.---}For $n\geq1$, consider a generic density operator in $\mathrm{span}\{\ket0,\ket n\}$,
\begin{equation}
 \rho_n(t,s)=(1-t)\, \ketbra00+t\, \ketbra nn
 +s\, (\ketbra0n+\ketbra n0),
 \label{eq:state}
\end{equation}
where $0\leq t\leq1$ and a phase rotation makes $s$ real and non-negative without loss of generality. Positive semi-definiteness of $\rho_n(t,s)$ amounts to $s^2\leq t(1-t)$.
Setting $\kappa(t,s)=(s^2+t^2)/t$, this is equivalent to $\kappa(t,s)\leq1$, and the state $\rho_n(t,s)$ admits the convex decomposition
\begin{equation}
 \rho_n(t,s)=(1-\kappa)\, \ketbra00
 +\kappa\, \ketbra{\varphi_n(t,s)}{\varphi_n(t,s)},
 \label{eq:pure-vacuum-mixture}
\end{equation}
where $\ket{\varphi_n(t,s)}=(s\ket0+t\ket n)/\sqrt{s^2+t^2}$. For nonzero $t$, the pure state $\ket{\varphi_n(t,s)}$ is non-Gaussian with bounded support over the Fock basis and therefore Wigner-negative in a bounded region of phase-space \cite[Theorem 1]{abreu2025inverse}. Hence, for small enough $t$, mixing it with the vacuum can make the resulting state $\rho_n(t,s)$ Wigner-positive. For each such $t$, we denote by $s_n(t)$ the maximal value of the Fock-coherence $s$ such that $\rho_n(t,s)$ remains Wigner-positive, and by $h_n(t)$ the corresponding Wigner entropy. We show in Appendix~\ref{app:0n} that, for small $t$, we have $s_n(t)=O(\sqrt t)$, and that the Wigner entropy behaves as
\begin{equation}
 h_n(t)=1+\ln\pi+(2n-2^n)t+o(t).
 \label{eq:main-expansion}
\end{equation}
Hence, for every fixed $n\geq3$, there exists $t_n>0$ such that $h_n(t)<1+\ln\pi$ whenever $0<t<t_n$. Note that Eq.~(\ref{eq:main-expansion}) does not lead to arbitrarily large violations of the conjecture for large $n$ because the threshold value~$t_n$ decreases very fast with $n$. In particular, numerical simulations give a maximal violation of at most $10^{-4}$ at $n=4$ for this family of counterexamples (see Figure~\ref{fig:results}). Note that choosing explicitly $s=\sqrt{t(1-t)(1-e\, n \, t^{1/n})}$ gives rise to counterexamples that do not reach the Wigner-positive boundary but exhibit the same behavior as Eq.~\eqref{eq:main-expansion}, see Appendix~\ref{app:0n}.
Note also that, letting $t\to 0$, we obtain counterexamples lying arbitrarily close to the vacuum.

\medskip

\emph{Physical interpretation.---}Equation~(\ref{eq:main-expansion}) has a clear interpretation in terms of the non-Gaussianity measure for a Wigner-positive state $\rho$, defined as  \cite{pizzimenti2023}
\begin{equation}
 D(W_\rho\Vert W_{\rho_G})=h(W_{\rho_G})-h(W_\rho),
 \label{eq:relative-wigner-entropy}
\end{equation}
that is, the relative Wigner entropy between the state $\rho$ and its Gaussification $\rho_G$ (i.e., the Gaussian state with the same first and second moments as $\rho$).
For $\rho=\rho_n(t,s_n(t))$, we have $h(W_\rho)=h_n(t)$, while its Gaussification $\rho_G$ is a thermal state with mean photon number $nt$, whose Wigner entropy is $h(W_{\rho_G})=1+\ln \pi + \ln(1+2nt)$. Hence, recalling $h(W_{\ket0\!\bra0})=1+\ln\pi$, Eq.~\eqref{eq:main-expansion} amounts to writing $h_n(t)=h(W_{\ket0\!\bra0})+G-D$, where, for small $t$, 
\begin{equation}
    \begin{aligned}
&G\equiv  h(W_{\rho_G})-h(W_{\ket0\!\bra0})=2nt+o(t),\\
&D\equiv D(W_{\rho}\Vert W_{\rho_G})=2^nt+o(t).
 \label{eq:relative-wigner-binary}
    \end{aligned}
\end{equation}
Thus, the uncertainty of $\rho$ as measured by $G$ (i.e., the Wigner entropy excess of the Gaussification of $\rho$ relative to the vacuum) contributes to a term of order $2nt$, whereas the non-Gaussianity of $\rho$ as measured by~$D$ (i.e., the Wigner relative entropy) leads to a negative contribution of order $2^n t$. The violation of the conjecture therefore results from the fact that the non-Gaussianity~$D$ exceeds the uncertainty $G$. The first violation at $n=3$ is precisely the point where $2^n$ exceeds $2n$.

Intuitively, one might expect the allowed non-Gaussianity $D$ of a Wigner-positive state to be constrained by its uncertainty $G$ (or its mixedness).
For pure states, $D$ is indeed fully constrained by Hudson's theorem (it must be zero). For increasingly mixed states, it is anticipated---according to the Wigner entropy conjecture---that $D$ cannot be large enough to allow for sub-vacuum entropy. However, the violating states $\rho_n(t,s)$ contradict this intuition. They have the notable property that $x$ and $p$ are uncorrelated (their covariance is zero) although they are not independent, as the mutual information $I(x{:}p)>0$. The latter is large enough so that $h(W_\rho)=h(P_x)+h(P_p)-I(x{:}p)<1+\ln\pi$ even though $h(P_x)+h(P_p)\ge 1+\ln\pi$.

\medskip

\emph{Proof of the conjecture in $\mathrm{span}\{\ket0,\ket2\}$.---}The previous counterexamples belong to $\mathrm{span}\{\ket0,\ket n\}$ for $n\ge3$, which begs the question of the existence of similar counterexamples for $n<3$. Since it is known that the Wigner entropy conjecture holds in $\mathrm{span}\{\ket0,\ket1\}$~\cite{qian2024}, we shall resolve this question for $n=2$. 
Note first that the state $\rho_2(t,s)$ is Wigner-negative whenever $t>\frac12$. For $0\leq t\leq\frac12$, a direct optimization yields the maximal value of the Fock-coherence $s$ such that $\rho_2(t,s)$ is Wigner-positive, namely
\begin{equation}
 s_2(t)=\sqrt t-\sqrt2\,t.
 \label{eq:s2}
\end{equation}
Set $t'=\sqrt{2t}$ and $\ket{\chi_\pm(t')}=\sqrt{1-t'}\ket0\pm\sqrt{t'}\ket1$.
By construction, the reduced output of a balanced beam-splitter fed with $\ket{\chi_+(t')}\otimes\ket{\chi_-(t')}$ is equal to $\rho_2(t,s_2(t))$, up to a phase shift, so $\rho_2(t,s_2(t))$ is a beam-splitter state [and similarly for $\rho_2(t,-s_2(t))$]. Moreover, for $0<t<\frac12$ and $s\le s_2(t)$, setting $\lambda=\frac12(1+s / s_2(t))$, we obtain
\begin{equation}
 \rho_2(t,s)=\lambda\, \rho_2(t,s_2(t))+(1-\lambda)\, \rho_2(t,-s_2(t)).
 \label{eq:n2-mixture}
\end{equation}
Hence, every Wigner-positive state in $\mathrm{span}\{\ket0,\ket2\}$ is a beam-splitter state and must consequently satisfy the Wigner entropy conjecture. 

Note that a similar reduction argument works for $\mathrm{span}\{\ket0,\ket1\}$. Indeed, any Wigner-positive state $\rho_1(t,s)$ can be written as a beam-splitter state obtained by mixing the state $\rho_1(2t,\sqrt2s)$ with the vacuum. This gives an alternative proof of the validity of the Wigner entropy conjecture in $\mathrm{span}\{\ket0,\ket1\}$~\cite{qian2024}. 

\medskip

\emph{Counterexamples in $\mathrm{span}\{\ket0,\ket1,\ket2\}$.---}We have seen that all Wigner-positive states in $\mathrm{span}\{\ket0,\ket1\}$ and $\mathrm{span}\{\ket0,\ket2\}$ satisfy the Wigner entropy conjecture, while there are counterexamples in $\mathrm{span}\{\ket0,\ket n\}$ for all $n\ge3$. We now resolve the status of the conjecture in $\mathrm{span}\{\ket0,\ket1,\ket2\}$ by constructing another family of counterexamples. Here, we consider  the state
\begin{equation}
    \ket{\phi_\epsilon}=\frac{\ket0+\sqrt{2\epsilon}\ket1
 +\epsilon\ket2/\sqrt2}{\sqrt{N_\epsilon}},
 \label{eq:three-level-state}
\end{equation}
for $\epsilon>0$, where $N_\epsilon=1+2\epsilon+\epsilon^2/2$ is a normalizing factor. This time, we mix it with another non-Gaussian pure state, namely the Fock state $\ket2$. The real part of the complex Wigner entropy of $\ket{\phi_\epsilon}$ is lower than $1+\ln\pi$, for small $\epsilon$, while that of $\ket2$ is greater. Here, both states are Wigner-negative, but in different regions of phase space, so that specific mixtures of them can be made Wigner-positive. Indeed, we show in Appendix~\ref{app:012} that the state
\begin{equation}
 \sigma_\epsilon=(1-\epsilon^4)\ket{\phi_\epsilon}\!\bra{\phi_\epsilon}+\epsilon^4\ket2\!\bra2
\end{equation}
is Wigner-positive for small enough $\epsilon$, while it satisfies
\begin{equation}
 h(W_{\sigma_\epsilon})=1+\ln\pi-\frac23\epsilon^3
 +O(\epsilon^{4}).
 \label{eq:three-level-entropy}
\end{equation}
Hence, the state $\sigma_\epsilon$ exhibits a sub-vacuum Wigner entropy for small $\epsilon$.

\medskip

\emph{Wigner--R\'enyi entropy.---}The Wigner--R\'enyi entropy of order $\alpha$ [see Eq.~(\ref{eq:WRentropy})] is $h_\alpha(W_{\ket0\!\bra0})=\ln\pi+\frac{\ln\alpha}{\alpha-1}$ for the vacuum. The Wigner--Rényi entropy conjecture states that the Wigner--Rényi entropy of order $\alpha$ of any Wigner-positive state is at least $h_\alpha(W_{\ket0\!\bra0})$~\cite{vanherstraeten2021}. The conjecture has been proven for all $\alpha\ge2$ \cite{diasprata2023}. For $0<\alpha<2$, we show that the conjecture does not hold, as our counterexamples to the Wigner entropy conjecture also provide counterexamples to the Wigner--Rényi entropy conjecture.

For every fixed $n\geq3$ and $0<\alpha<2$, denoting again $\rho=\rho_n(t,s_n(t))$, we prove in Appendix~\ref{app:Renyi} that for small $t>0$:
\begin{align}
 h_\alpha(W_\rho)
 &=h_\alpha(W_{\ket0\!\bra0})+C_{n,\alpha}t+o(t),
 \label{eq:binary-renyi}
\end{align}
 where $C_{n,\alpha}=-2\sum_{k=1}^{n-2}\left[\binom{n-1}{k}-1\right]\left(\frac{2-\alpha}{\alpha}\right)^k<0$. Similarly, for small $\epsilon>0$, we find
\begin{equation}
 h_\alpha(W_{\sigma_\epsilon})
 =h_\alpha(W_{\ket0\!\bra0})-\frac{2(2-\alpha)}{3\alpha}\epsilon^3+O(\epsilon^4).
 \label{eq:three-level-renyi}
\end{equation}
Both counterexample families therefore disprove the Wigner--R\'enyi conjecture throughout the range $0<\alpha<2$.

\medskip

\emph{Discussion and outlook.---}We have shown that the Wigner entropy conjecture, the Wigner--Rényi entropy conjecture for all $0<\alpha<2$, and the Wigner majorization conjecture are all false in general. This is somewhat surprising, given that these conjectures not only appear to be natural generalizations of the entropic uncertainty principle, but also hold true for broad classes of Wigner-positive states. As a matter of fact, while conceptually simple, our minimal counterexamples must necessarily have high purity, nonzero coherence, and lie outside of the large class of beam-splitter states. 

Interestingly, our construction of counterexamples rests upon the existence of pure non-Gaussian states for which the real part of the complex-valued Wigner entropy is lower than that of the vacuum. In our counterexamples, the non-Gaussianity defined in \cite{pizzimenti2023} outweighs the Gaussian quadrature uncertainty, defined by the Wigner entropy of the Gaussification. This motivates further characterization of all possible counterexamples and further study of the complex-valued Wigner entropy, which we leave to future work.
 
Regarding Wigner-positive states, an analytical lower bound on the Wigner entropy is $\ln(2\pi)$ \cite{diasprata2023}. Our results show that the best lower bound lies strictly below $1+\ln\pi$, while a quick numerical search only identified violations of the Wigner entropy conjecture of order at most $10^{-3}$. The small magnitude of these violations may explain why counterexamples have been overlooked in previous works.  A natural follow-up question is to determine the best possible lower bound and whether this lower bound is attained or is an infimum (note the significant gap of $\ln(e/2)$ with the known lower bound \cite{diasprata2023}).

Finally, we note that the observation of a state with sub-vacuum Wigner entropy necessarily excludes broad classes of Wigner-positive states, so our work may have interesting prospects in the context of state certification.

\medskip

\textit{Acknowledgments.---}We acknowledge interesting discussions with J.\ N.\ Prata, N.\ Dias and J.\ Davis and thank J.\ N.\ Prata and R.\ Wagner for their feedback.
Z.V.H.\ and U.C.\ acknowledge funding from the European Union's Horizon Europe Framework Programme through the EIC Pathfinder Challenge project VeriQuB under Grant Agreement No.~101114899.
N.J.C.\ acknowledges support from the Fonds de la Recherche Scientifique–FNRS (Belgium) under Grant No.~T.0060.26 as well as project CHEQS
within the Excellence of Science (EOS) program.

\medskip

\textit{Disclaimer.---}The authors acknowledge the use of a generative AI tool in this work and take full responsibility for the content of this manuscript.

\medskip

\textit{Note added.---}While finalizing this manuscript, we became aware of a recent preprint introducing similar counterexamples \cite{he2026}. 

\bibliography{bib}

\onecolumngrid
\bigskip
\bigskip
\hrule
\clearpage
\appendix
\setcounter{secnumdepth}{2}

\begin{center}
    {\Huge Appendices}
\end{center}

These appendices contain the proofs of the results stated in the main text.
Appendix~\ref{app:prelim} fixes our conventions and collects a few elementary facts used throughout; Appendix~\ref{app:0n} treats the states supported on $\{\ket0,\ket n\}$, including the cases $n=1,2$ for which the conjecture holds; Appendix~\ref{app:012} treats the family $\sigma_\epsilon$; and Appendix~\ref{app:Renyi} extends both constructions to Wigner--R\'enyi entropies of order $0<\alpha<2$.

\section{Preliminary material}
\label{app:prelim}

\subsection{Conventions and coordinates}

We use $\hbar=1$, natural logarithms, and the convention $0\ln0=0$.
The number operator is $\hat n=(\hat x^2+\hat p^2-1)/2$, and the vacuum Wigner function is denoted by
\begin{equation}
 W_0(x,p):=W_{\ketbra00}(x,p)=\frac1\pi e^{-(x^2+p^2)}.
 \label{eq:s-vacuum-wigner}
\end{equation}
All the states considered below have finite support in the Fock basis, so that their Wigner functions are the product of a Gaussian function and a polynomial function; in particular, all the Gaussian moments appearing in our calculations are finite.
Recall that the Wigner transform is linear, satisfies $\int_{\R^2}W_A\dd x\dd p=\Tr A$ for the finite-rank operators considered here, and obeys the overlap identity~\cite{moyal1949}
\begin{equation}
 \Tr(AB)=2\pi\int_{\R^2}W_AW_B\dd x\dd p.
 \label{eq:s-moyal}
\end{equation}

For polar integration, we write
\begin{equation}
 x+ip=\sqrt R\,e^{i\theta},\qquad R=x^2+p^2,\qquad
 \dd x\dd p=\frac12\dd R\dd\theta.
 \label{eq:s-polar-conventions}
\end{equation}
Thus $R$ is the squared radius, and $W_0\dd x\dd p=e^{-R}\dd R\dd\theta/(2\pi)$.

\subsection{Fock basis and Laguerre polynomials}

For integers $0\le m\le n$, the Wigner transform of a Fock basis operator is~\cite{wunsche1998laguerre}
\begin{equation}
 \frac{W_{\ketbra mn}(R,\theta)}{W_0(R)}
 =(-1)^m\sqrt{\frac{m!}{n!}}
 (\sqrt{2R}\,e^{i\theta})^{n-m}L_m^{(n-m)}(2R).
 \label{eq:s-associated-laguerre}
\end{equation}
Here $L_m^{(n-m)}$ is a generalized Laguerre polynomial; the case $m>n$ follows from $W_{\ketbra nm}=W_{\ketbra mn}^*$.
In particular, for the operators $\ketbra{n}{n}$ and $\ketbra{0}{n}+\ketbra{n}{0}$, we obtain
\begin{align}
 W_{\ketbra nn}(R)&=W_0(R)(-1)^nL_n(2R),
 \label{eq:s-diagonal-fock}\\
 W_{\ketbra0n+\ketbra n0}(R,\theta)
 &=W_0(R)\frac{2(2R)^{n/2}}{\sqrt{n!}}\cos(n\theta).
 \label{eq:s-interference}
\end{align}
We use the convention~\cite{olver2010}
\begin{equation}
 L_n(z)=\sum_{k=0}^n(-1)^k\binom nk\frac{z^k}{k!},
 \qquad
 (-1)^nL_n(z)=\frac{z^n-n^2z^{n-1}}{n!}+O(z^{n-2})
 \quad(z\to\infty,\ n\ge2).
 \label{eq:s-laguerre-full}
\end{equation}
The $n$ zeros $r_1,\ldots,r_n$ of $L_n$ are simple and strictly positive~\cite[Sec.~18.2(vi)]{olver2010}, and
\begin{equation}
 (-1)^nL_n(z)=\frac1{n!}\prod_{j=1}^n(z-r_j),
 \qquad \sum_{j=1}^n r_j=n^2.
 \label{eq:s-laguerre-zeros}
\end{equation}
Indeed, the coefficient of $z^{n-1}$ in this product is $-\sum_{j=1}^n r_j/n!$, which must match the coefficient $-n^2/n!$ in Eq.~\eqref{eq:s-laguerre-full}.
These facts will be used in Appendix~\ref{app:explicit-coherence}.

\subsection{Gaussian integrals}

We use the following notation for integration with a normalized Gaussian weight:
\begin{equation}
 \int_{\R^2}f\dd\nu_\alpha
 :=\frac{\alpha}{\pi}\int_{\R^2}f(x,p)e^{-\alpha R}\dd x\dd p,
 \qquad \alpha>0.
 \label{eq:s-nu-alpha}
\end{equation}
Thus $\nu_\alpha$ is a probability measure, with $\dd\nu_1=W_0\dd x\dd p$ for the vacuum.
The weights $\nu_{\alpha}$ with $\alpha\neq 1$ will be needed for the R\'enyi entropies, since $W^{\alpha}_{0}$ is proportional to $e^{-\alpha R}$.
For non-negative integers $j,\ell,k$, the required integrals are
\begin{align}
 \int_0^\infty e^{-\alpha R}R^k\dd R&=\frac{k!}{\alpha^{k+1}},
 \label{eq:s-gaussian-radial}\\
 \int_{\R^2}x^{2j}p^{2\ell}\dd\nu_\alpha
 &=\frac{(2j-1)!!(2\ell-1)!!}{(2\alpha)^{j+\ell}}.
 \label{eq:s-three-renyi-moments}
\end{align}
Here $(2j-1)!!=1\cdot3\cdots(2j-1)$ for $j\ge1$ is the double factorial, with the convention $(-1)!!=1$.
Any moment odd in $x$ or $p$ vanishes.
For $n\ge1$, the angular integrals are
\begin{align}
 \int_0^{2\pi}\cos(n\theta)\dd\theta=0,
 \qquad \int_0^{2\pi}\cos^2(n\theta)\dd\theta=\pi.
\end{align}
Integrating the finite sum in Eq.~\eqref{eq:s-laguerre-full} gives
\begin{equation}
 \alpha\int_0^\infty e^{-\alpha R}(-1)^nL_n(2R)\dd R
 =\left(\frac{2-\alpha}{\alpha}\right)^n.
 \label{eq:s-renyi-laguerre-moment}
\end{equation}
At $\alpha=1$, this reduces to the normalization $\int_0^\infty e^{-R}(-1)^nL_n(2R)\dd R=1$.

\subsection{Entropies relative to the vacuum}

For a normalized non-negative Wigner function, the Wigner entropy and Wigner--R\'enyi entropy of order $\alpha$ are defined as~\cite{renyi1961,vanherstraeten2021}
\begin{equation}
 h(W)=-\int_{\R^2}W\ln W\dd x\dd p,
 \qquad
 h_\alpha(W)=\frac1{1-\alpha}\ln\int_{\R^2}W^\alpha\dd x\dd p
 \quad(\alpha>0,\ \alpha\ne1).
 \label{eq:s-entropy-definitions}
\end{equation}
For the vacuum, these definitions give
\begin{equation}
 h(W_0)=1+\ln\pi,
 \qquad h_\alpha(W_0)=\ln\pi+\frac{\ln\alpha}{\alpha-1}.
 \label{eq:s-vacuum-renyi}
\end{equation}
For normalized $W$, the Wigner--R\'enyi entropy tends to the Wigner entropy as $\alpha\to1$.

It is convenient to express a Wigner function as a relative perturbation of the vacuum,
\begin{equation}
 W=W_0(1+\delta),\qquad 1+\delta\ge0,
 \qquad \int_{\R^2}\delta\dd\nu_1=0.
 \label{eq:s-relative-perturbation}
\end{equation}
Its relative entropy with respect to the vacuum is
\begin{equation}
 D(W\Vert W_0)
 \coloneqq
\int_{\mathbb{R}^2}
W\ln\frac{W}{W_0}\dd x\dd p
 =
 \int_{\R^2}(1+\delta)\ln(1+\delta)\dd\nu_1.
 \label{eq:s-relative-entropy}
\end{equation}
Since $\ln W_0=-\ln\pi-R$ and $\int RW\dd x\dd p=2\langle\hat n\rangle+1$, the entropy difference is
\begin{equation}
 h(W)-h(W_0)=2\langle\hat n\rangle-D(W\Vert W_0),
 \label{eq:s-entropy-identity}
\end{equation}
where $\langle\hat{n}\rangle=\Tr(\rho\,\hat{n})$ is the mean photon number of the state.
Thus the entropy lies below the vacuum value precisely when $D(W\Vert W_0)>2\langle\hat n\rangle$.
For the Wigner--R\'enyi entropies, the corresponding identity is
\begin{equation}
 h_\alpha(W)-h_\alpha(W_0)
 =\frac1{1-\alpha}\ln\int_{\R^2}(1+\delta)^\alpha\dd\nu_\alpha.
 \label{eq:s-renyi-ratio}
\end{equation}

\subsection{Control of entropy expansions}
\label{sec:s-expansion-control}

All the counterexamples constructed in this work are states close to the vacuum, and our strategy to compute their entropy is always the same: we write $W=W_0(1+\delta)$ as in Eq.~\eqref{eq:s-relative-perturbation}, where $\delta$ vanishes as the state approaches the vacuum, and we expand the integrals appearing in Eqs.~\eqref{eq:s-relative-entropy} and~\eqref{eq:s-renyi-ratio} in powers of $\delta$.
Note that $\delta$ is small only in a pointwise sense: for the states considered here, $W/W_0$ is a polynomial in $x$ and $p$, so that $\delta$ tends to zero at each fixed point but grows with $R$, and $\delta$ even reaches the value $-1$ at the zeros of the Wigner function.
The Taylor expansions of $(1+z)\ln(1+z)$ and $(1+z)^\alpha$ around $z=0$ can therefore not be integrated term by term without further justification.
More precisely, we shall need remainder bounds valid on the whole range $z\ge-1$, which we collect here since they are used repeatedly in Appendices~\ref{app:0n}--\ref{app:Renyi}.

The first bound concerns the Shannon relative entropy~\eqref{eq:s-relative-entropy}.
For every $z\ge-1$, we have
\begin{equation}
 0\le\Phi(z):=(1+z)\ln(1+z)-z\le z^2,
 \qquad
 \lim_{z\to0}\frac{\Phi(z)}{z^2}=\frac12.
 \label{eq:s-shannon-global-bound}
\end{equation}
The lower bound follows from convexity and $\Phi(0)=\Phi'(0)=0$, the upper bound from $\ln(1+z)\le z$ (the case $z=-1$ follows by continuity), and the limit from the Taylor expansion at zero.
Its R\'enyi analogue, which concerns the integrand of Eq.~\eqref{eq:s-renyi-ratio}, reads, for fixed $0<\alpha<2$ and $z\ge-1$,
\begin{equation}
 |(1+z)^\alpha-1-\alpha z|\le C_\alpha z^2,
 \qquad
 \lim_{z\to0}\frac{(1+z)^\alpha-1-\alpha z}{z^2}
 =\frac{\alpha(\alpha-1)}2,
 \label{eq:s-renyi-linear-bound}
\end{equation}
where $C_\alpha$ is a finite constant depending only on $\alpha$.
Indeed, the quotient extends continuously to $z=0$ and $z=-1$ and tends to zero at infinity because $\alpha<2$, so that it is bounded on $[-1,\infty)$.

In several of the cases treated below, the leading term of this expansion cancels out, and the sign of the entropy difference is only decided at higher order.
We shall then need to carry the expansion further, while keeping its remainder under control on the whole range $z\ge-1$.
For $F(y)=y\ln y$ or $F(y)=y^\alpha$ with fixed $0<\alpha<2$, we shall therefore also use the bound
\begin{equation}
 \left|F(1+z)-\sum_{k=0}^7\frac{F^{(k)}(1)}{k!}z^k\right|
 \le C_F|z|^8,
 \qquad z\ge-1,
 \label{eq:s-global-taylor}
\end{equation}
where $C_F$ is a finite constant depending only on $F$ (and hence possibly on $\alpha$).
To prove it, divide the left-hand side by $|z|^8$: the resulting quotient is bounded near zero by Taylor's theorem, remains finite at $z=-1$ by continuity, and tends to zero at infinity because $F(1+z)=o(z^8)$ while the Taylor polynomial has degree at most seven.
Since it is continuous elsewhere, it is bounded on $[-1,\infty)$, which proves Eq.~\eqref{eq:s-global-taylor}.

\section{States supported on the vacuum and one Fock level}
\label{app:0n}

For $n\ge1$, consider again the family of states of Eq.~\eqref{eq:state} in the main text,
\begin{equation}
 \rho_n(t,s)=(1-t)\ketbra00+t\ketbra nn
 +s(\ketbra0n+\ketbra n0),\qquad 0\le t\le1.
 \label{eq:s-state}
\end{equation}
All asymptotic statements for this family refer to $t\to0^+$ with $n$ fixed; constants in the remainder estimates may depend on $n$.
A phase rotation makes $s$ real and non-negative without changing either Wigner positivity or entropy.
Positive semi-definiteness of $\rho_n(t,s)$ amounts to $s^2\le t(1-t)$.
We first determine the largest coherence allowed by Wigner positivity, then give an explicit alternative that produces the same leading entropy decrease.

\subsection{Wigner positivity and maximal coherence}
\label{sec:s-maximal-coherence}

Equations~\eqref{eq:s-diagonal-fock} and \eqref{eq:s-interference} give
\begin{equation}
 \frac{W_{\rho_n(t,s)}}{W_0}
 =1-t+t(-1)^nL_n(2R)
 +\frac{2s(2R)^{n/2}}{\sqrt{n!}}\cos(n\theta).
 \label{eq:s-family-wigner}
\end{equation}
For each $R>0$, the minimum over the angle is obtained at $\cos(n\theta)=-1$.
Consequently,
\begin{equation}
 W_{\rho_n(t,s)}\ge0
 \quad\Longleftrightarrow\quad
 1-t+t(-1)^nL_n(2R)\ge\frac{2s(2R)^{n/2}}{\sqrt{n!}}
 \quad\text{for all }R\ge0.
 \label{eq:s-radial-positivity}
\end{equation}
As shown in Appendix~\ref{app:explicit-coherence}, this condition is satisfied by a whole interval of values $s\geq 0$ as soon as $t$ is small enough.

At fixed $t$, the states $\rho_n(t,s)$ and $\rho_n(t,-s)$ have the same Wigner entropy, since their Wigner functions are related by a phase-space rotation.
Mixing these two states produces all possible coherence values between $-s$ and $s$ and cannot lower their common Wigner entropy, by concavity of the Shannon entropy.
We therefore seek the largest non-negative coherence compatible with Wigner positivity, which minimizes the Wigner entropy within this family at fixed $t$.

We must also check that this choice satisfies the positive semi-definiteness constraint $s\le\sqrt{t(1-t)}$.
At equality, $\rho_n(t,s)$ is the projector onto $\sqrt{1-t}\ket0+\sqrt t\ket n$.
This pure state is non-Gaussian, so its Wigner function takes negative values by Hudson's theorem~\cite{hudson1974}.
Since the right-hand side of the radial condition increases with $s$, Wigner positivity imposes a strictly smaller upper bound and therefore already enforces the positive semi-definiteness constraint.
Taking the infimum of the bounds in Eq.~\eqref{eq:s-radial-positivity} over $R>0$ gives the maximal admissible coherence:
\begin{equation}
 s_n(t)=
 \inf_{R>0}\frac{\sqrt{n!}[1-t+t(-1)^nL_n(2R)]}{2(2R)^{n/2}}.
 \label{eq:s-sn}
\end{equation}

Following the main text, we write $h_n(t):=h(W_{\rho_n(t,s_n(t))})$ for the Wigner entropy at maximal coherence.

\subsection{An explicit admissible coherence}
\label{app:explicit-coherence}

The expression for $s_n(t)$ in Eq.~\eqref{eq:s-sn} requires a minimization over $R$.
We now seek a simpler explicit expression for an admissible coherence with the same leading behavior as $s_n(t)$ when $t\to0^+$.
This will also provide an explicit lower bound on $s_n(t)$.

We consider $0\le s\le\sqrt{t(1-t)}$ and write $z=2R$.
The Wigner positivity condition~\eqref{eq:s-radial-positivity} becomes
\begin{align}
 1-t+t(-1)^nL_n(z)
 \ge \frac{2s z^{n/2}}{\sqrt{n!}}.
\end{align}
We split the domain at $z=n^2$, since every Laguerre zero satisfies $0<r_j\le\sum_{k=1}^n r_k=n^2$.
On $0\le z\le n^2$, the left-hand side tends uniformly to $1$ as $t\to0^+$, because $L_n(z)$ is bounded.
The right-hand side tends uniformly to $0$ for all allowed $s$, because $z^{n/2}$ is bounded and $s\le\sqrt t$.
Thus the condition holds throughout this interval for all sufficiently small $t>0$.

For $z\ge n^2$, the factorization in Eq.~\eqref{eq:s-laguerre-zeros} gives
\begin{equation}
 (-1)^nL_n(z)
 =\frac{z^n}{n!}\prod_{j=1}^n\left(1-\frac{r_j}{z}\right)
 \ge\frac{z^n-n^2z^{n-1}}{n!}.
 \label{eq:s-laguerre-tail-bound}
\end{equation}
Here $0\le r_j/z\le1$, so the inequality follows from $\prod_j(1-a_j)\ge1-\sum_j a_j$ for $a_j\in[0,1]$.

We divide the Wigner positivity condition by $1-t$ and introduce $u$ so that the leading term $tz^n/[(1-t)n!]$ becomes $u^n$:
\begin{align}
 u=z\left(\frac{t}{(1-t)n!}\right)^{1/n},
 \qquad
 \gamma=n^2\left(\frac{t}{(1-t)n!}\right)^{1/n}.
\end{align}
Using Eq.~\eqref{eq:s-laguerre-tail-bound}, we obtain
\begin{align}
 \frac{1-t+t(-1)^nL_n(z)}{1-t}
 &\ge 1+u^n-\gamma u^{n-1},\\
 \frac{2s z^{n/2}}{(1-t)\sqrt{n!}}
 &=\frac{2s}{\sqrt{t(1-t)}}\,u^{n/2}.
\end{align}
We therefore seek a lower bound on $1+u^n-\gamma u^{n-1}$ proportional to $u^{n/2}$.

For sufficiently small $t>0$, we have $0<\gamma<1$.
Using $u^{n-1}\le1$ when $0\le u\le1$ and $u^{n-1}\le u^n$ when $u\ge1$ gives
\begin{align}
 1+u^n-\gamma u^{n-1}
 \ge
 \begin{cases}
  (1-\gamma)+u^n, & 0\le u\le1,\\
  1+(1-\gamma)u^n, & u\ge1.
 \end{cases}
\end{align}
By the arithmetic--geometric mean inequality, both lower bounds are at least $2\sqrt{1-\gamma}\,u^{n/2}$.
Comparison with the normalized right-hand side therefore shows that Wigner positivity is guaranteed whenever
\begin{align}
 0\le s\le\sqrt{t(1-t)(1-\gamma)}.
\end{align}

This already provides a lower bound on $s_n(t)$.
We simplify it by replacing $\gamma$ with an upper bound that does not involve $n!$.
Indeed, $\ln(n!)\ge\int_1^n\ln x\,\dd x>n\ln n-n$ implies $(n!)^{1/n}>n/e$, and hence
\[
 \frac{\gamma}{t^{1/n}}
 \longrightarrow\frac{n^2}{(n!)^{1/n}}<en.
\]
Thus $\gamma\le en\,t^{1/n}<1$ for all sufficiently small $t>0$.
Replacing $\gamma$ by this upper bound gives a smaller admissible coherence, and therefore
\begin{equation}
 \sqrt{(1-t)(1-en\,t^{1/n})}
 \le\frac{s_n(t)}{\sqrt t}
 \le\sqrt{1-t}.
 \label{eq:s-coherence-limit}
\end{equation}
Both bounds tend to one when $t\to0^+$, proving $s_n(t)^2=t+o(t)$, and implying $s_n(t)=O(\sqrt{t})$.
For $n=1,2$, the same asymptotic behavior follows from the bounds in Appendix~\ref{app:s-low-levels}.

For $n\ge3$, the construction yields the following explicit admissible coherence, with the same leading behavior as $s_n(t)$:
\begin{equation}
 \widetilde s_n(t)=\sqrt{t(1-t)\bigl(1-en\,t^{1/n}\bigr)}.
 \label{eq:s-explicit-coherence}
\end{equation}

\subsection{Asymptotic Wigner entropy}
\label{sec:s-binary-entropy}

We now take either the maximal coherence $s=s_n(t)$, for any fixed $n\ge1$, or the explicit coherence $s=\widetilde s_n(t)$, for fixed $n\ge3$.
Both choices give Wigner-positive states for sufficiently small $t$, with $s^2=t+o(t)$.
The coefficient of $\ket n$ contributes $2nt$ to Eq.~\eqref{eq:s-entropy-identity}, while the entropy-lowering contribution will come from the square of the coherence.

For use here and in the Wigner--R\'enyi entropy calculations in Appendix \ref{app:Renyi}, write
\begin{equation}
 A_n(R)=(-1)^nL_n(2R)-1,
 \qquad
 B_n(R,\theta)=\frac{2(2R)^{n/2}}{\sqrt{n!}}\cos(n\theta),
 \label{eq:s-binary-components}
\end{equation}
so that the relative perturbation is
\begin{equation}
 \delta_{n,t}=\frac{W_{\rho_n(t,s)}}{W_0}-1=tA_n+sB_n.
 \label{eq:s-relative-deviation}
\end{equation}
Equations~\eqref{eq:s-gaussian-radial} and \eqref{eq:s-renyi-laguerre-moment}, together with integration over $\theta$, give
\begin{equation}
 \int A_n\dd\nu_1=\int B_n\dd\nu_1=0,
 \qquad
 \int B_n^2\dd\nu_\alpha
 =\frac{4\cdot2^n}{n!}\cdot\frac{\alpha}{2\pi}
 \int_0^\infty e^{-\alpha R}R^n\dd R\int_0^{2\pi}\cos^2(n\theta)\dd\theta
 =\frac{2^{n+1}}{\alpha^n}.
 \label{eq:s-binary-moments}
\end{equation}
In particular, $\int B_n^2\dd\nu_1=2^{n+1}$, which is the value needed here. We use the case of general $\alpha$ in Appendix~\ref{app:Renyi}.

We now expand $D(W_{\rho_n(t,s)}\Vert W_0)$ to first order in $t$.
Since $\int\delta_{n,t}\dd\nu_1=0$, Eq.~\eqref{eq:s-relative-entropy} gives $D(W_{\rho_n(t,s)}\Vert W_0)=\int\Phi(\delta_{n,t})\dd\nu_1$ with $\Phi$ as in Eq.~\eqref{eq:s-shannon-global-bound}.
At each fixed point, $\delta_{n,t}/\sqrt t\to B_n$ because $s/\sqrt t\to1$, hence $\Phi(\delta_{n,t})/t\to B_n^2/2$ by the limit in Eq.~\eqref{eq:s-shannon-global-bound}.
Moreover, $s^2\le t$ and $t\le1$ imply the bound
\begin{equation}
 0\le\frac{\Phi(\delta_{n,t})}{t}
 \le\frac{\delta_{n,t}^2}{t}\le2A_n^2+2B_n^2.
 \label{eq:s-binary-dominating-bound}
\end{equation}
The right-hand side is integrable against $\nu_1$ because it grows only polynomially with $R$.
Dominated convergence therefore gives
\begin{equation}
 D(W_{\rho_n(t,s)}\Vert W_0)
 =\int\Phi(\delta_{n,t})\dd\nu_1
 =2^nt+o(t).
 \label{eq:s-nonlinear-result}
\end{equation}
Note that the argument does not require $1+\delta_{n,t}$ to stay away from zero, since the bound~\eqref{eq:s-shannon-global-bound} holds down to $z=-1$; the points where the Wigner function of $\rho_n(t,s_n(t))$ vanishes therefore cause no difficulty.
Since $\langle\hat n\rangle=nt$, Eq.~\eqref{eq:s-entropy-identity} now yields
\begin{equation}
 h(W_{\rho_n(t,s)})=1+\ln\pi+(2n-2^n)t+o(t).
 \label{eq:s-main-expansion}
\end{equation}
In particular, this is the expansion of $h_n(t)$, and it also holds for the explicit admissible coherence.
For every fixed $n\ge3$, the coefficient is negative, so both choices give counterexamples for all sufficiently small positive $t$.
They approach the vacuum in trace distance: $\tfrac12\|\rho_n(t,s)-\ketbra00\|_1=\sqrt{t^2+s^2}\to0$.

Finally, let us justify the identification of the Gaussification of $\rho_n(t,s)$ used in the main text.
For $n\ge3$, the coherence between $\ket0$ and $\ket n$ contributes to neither the first nor the second moments, so the first moments vanish and the covariance matrix is $(nt+1/2)I$.
The Gaussian state with these moments is therefore the thermal state with mean photon number $nt$.

\subsection{The cases \texorpdfstring{$n=1$ and $n=2$}{n=1 and n=2}}
\label{app:s-low-levels}

Recall from the main text that a beam-splitter state is the reduced output of a balanced beam splitter fed with a separable two-mode input, and that all such states satisfy the Wigner entropy conjecture~\cite{interference2025}.
We show here that every Wigner-positive state in $\operatorname{span}\{\ket0,\ket1\}$ or $\operatorname{span}\{\ket0,\ket2\}$ is a beam-splitter state, which proves the conjecture in both sectors.
Throughout this subsection, the beam splitter acts as $a^\dagger\mapsto(a^\dagger+b^\dagger)/\sqrt2$, $b^\dagger\mapsto(a^\dagger-b^\dagger)/\sqrt2$, $\ket{nm}$ denotes $n$ photons in the first mode and $m$ in the second, and the second mode is the one being traced out.

For $n=1$ and $t>0$, completing the square gives
\begin{equation}
 \frac{W_{\rho_1(t,s)}}{W_0}
 =2t\left[\left(x+\frac{s}{\sqrt2t}\right)^2+p^2\right]
 +1-2t-\frac{s^2}{t}.
 \label{eq:s-n1-positivity}
\end{equation}
Wigner positivity is therefore equivalent to $t\le1/2$ and $s^2\le t(1-2t)$, with maximal coherence $s_1(t)=\sqrt{t(1-2t)}$, i.e., Eq.~\eqref{eq:s-sn} for $n=1$.
These are exactly the conditions for $\rho_1(2t,\sqrt2s)$ to be a valid input state.
A direct derivation shows that mixing that state with the vacuum at a balanced beam splitter gives $\rho_1(t,s)$.

For $n=2$, minimizing the ratio $W_{\rho_2(t,s)}/W_0$ over $\theta$ at fixed $R$ gives
\begin{align}
 \min_\theta\frac{W_{\rho_2(t,s)}(R,\theta)}{W_0(R)}
 =1-(4t+2\sqrt2s)R+2tR^2.
\end{align}
Minimizing this quadratic over $R\ge0$ yields
\begin{equation}
 0\le t\le\frac12,
 \qquad 0\le s\le s_2(t)=\sqrt t-\sqrt2t.
 \label{eq:s-s2}
\end{equation}
Thus $s_2(t)$ is the maximal coherence~\eqref{eq:s-sn} for $n=2$.
Now set $t'=\sqrt{2t}$ and consider the product input
\begin{align}
 \bigl(\sqrt{1-t'}\ket0+\sqrt{t'}\ket1\bigr)
 \otimes\bigl(\sqrt{1-t'}\ket0-\sqrt{t'}\ket1\bigr).
\end{align}
The beam splitter maps it to
\begin{equation}
 (1-t')\ket{00}+\sqrt{2t'(1-t')}\ket{01}
 -\frac{t'}{\sqrt2}\ket{20}+\frac{t'}{\sqrt2}\ket{02}.
 \label{eq:s-n2-beamsplitter-output}
\end{equation}
Tracing out the second mode gives $\rho_2(t,-s_2(t))$, and a phase rotation gives the opposite coherence.
For $0<t<1/2$ and $0\le s\le s_2(t)$, all intermediate coherences are obtained from
\begin{align}
 \rho_2(t,s)=\lambda\rho_2(t,s_2(t))+(1-\lambda)\rho_2(t,-s_2(t)),
 \qquad \lambda=\frac12\left(1+\frac{s}{s_2(t)}\right).
\end{align}
These mixtures remain beam-splitter states because separable inputs form a convex set.
The endpoints $t=0$ and $t=1/2$ follow directly from the construction with $t'=0$ and $t'=1$.
This proves the entropy bound in both sectors.

For completeness, we also determine the first nonzero term of $h_2(t)-h(W_0)$, where $h_2(t)=h(W_{\rho_2(t,s_2(t))})$, which shows that the $n=2$ curve in Fig.~\ref{fig:results} approaches the vacuum value from above:
\begin{equation}
 h_2(t)-h(W_0)=8\sqrt2\,t^{3/2}+O(t^2).
 \label{eq:s-n2-entropy-expansion}
\end{equation}
Indeed, write $\delta_{2,t}=tA_2+s_2(t)B_2$ as in Eq.~\eqref{eq:s-relative-deviation}.
Angular integration gives $\int A_2B_2\dd\nu_1=\int B_2^3\dd\nu_1=0$, and hence
\begin{align}
 \int\delta_{2,t}^2\dd\nu_1=8s_2(t)^2+O(t^2),
 \qquad \int\delta_{2,t}^3\dd\nu_1=O(t^2).
\end{align}
For $4\le k\le8$, the bound in Eq.~\eqref{eq:s-binary-dominating-bound} gives $\int|\delta_{2,t}|^k\dd\nu_1=O(t^{k/2})$.
The Taylor expansion with the bound~\eqref{eq:s-global-taylor} therefore yields $D(W_{\rho_2(t,s_2(t))}\Vert W_0)=4s_2(t)^2+O(t^2)$.
Using $h_2(t)-h(W_0)=4t-D(W_{\rho_2(t,s_2(t))}\Vert W_0)$ and substituting Eq.~\eqref{eq:s-s2} proves Eq.~\eqref{eq:s-n2-entropy-expansion}.

Finally, no Wigner-positive state is supported only on $\operatorname{span}\{\ket1,\ket2\}$: indeed, Eq.~\eqref{eq:s-moyal} gives $\langle0|\rho|0\rangle=2\pi\int W_\rho W_0\dd x\dd p>0$ for every Wigner-positive state, since $W_0$ is strictly positive.
Together with the two cases above, this shows that all three lowest Fock levels are needed for a counterexample in $\operatorname{span}\{\ket0,\ket1,\ket2\}$, which is the object of the next appendix.

\section{Counterexamples in the \texorpdfstring{$0$--$1$--$2$}{0--1--2} sector}
\label{app:012}

We now construct counterexamples supported on $\operatorname{span}\{\ket0,\ket1,\ket2\}$.
Consider again the normalized vector of Eq.~\eqref{eq:three-level-state} in the main text,
\begin{equation}
 \ket{\phi_\epsilon}
 =\frac{\ket0+\sqrt{2\epsilon}\ket1+\epsilon\ket2/\sqrt2}
 {\sqrt{N_\epsilon}},
 \qquad N_\epsilon=1+2\epsilon+\frac{\epsilon^2}{2}.
 \label{eq:s-three-pure-state}
\end{equation}
This choice of coefficients leads to a simple expression for the Wigner function used in the proof below.
We mix this state with $\ket2$:
\begin{equation}
 \sigma_\epsilon=(1-\epsilon^4)\ketbra{\phi_\epsilon}{\phi_\epsilon}
 +\epsilon^4\ketbra22,
 \qquad 0<\epsilon<1.
 \label{eq:s-three-mixture}
\end{equation}
The weight $\epsilon^4$ will suffice to restore Wigner positivity while leaving the polynomial ratio to $W_0$ unchanged through order $\epsilon^3$.
All asymptotic statements for this family refer to $\epsilon\to0^+$.
Note that $\sigma_\epsilon$ is a mixed state of rank two since $\ket{2}$ and $\ket{\phi_\epsilon}$ are linearly independent.
We first prove that $\sigma_\epsilon$ is Wigner-positive for sufficiently small $\epsilon>0$, then compute its entropy up to order $\epsilon^3$.

\subsection{Wigner positivity}

Expanding $\ketbra{\phi_\epsilon}{\phi_\epsilon}$ in the Fock basis and applying Eq.~\eqref{eq:s-associated-laguerre} gives
\begin{equation}
 \frac{W_{\ketbra{\phi_\epsilon}{\phi_\epsilon}}}{W_0}
 =\frac{(q_\epsilon-\epsilon)^2-\epsilon^2/2}{N_\epsilon},
 \qquad
 q_\epsilon=1+2\sqrt\epsilon\,x+\epsilon R
 =\epsilon\bigl[(x+\epsilon^{-1/2})^2+p^2\bigr].
 \label{eq:s-three-pure-wigner}
\end{equation}
Consequently,
\begin{equation}
 \frac{W_{\sigma_\epsilon}}{W_0}
 =\frac{1-\epsilon^4}{N_\epsilon}
 \bigl[(q_\epsilon-\epsilon)^2-\epsilon^2/2\bigr]
 +\epsilon^4 P_2(R),
 \qquad P_2(R)=\frac{W_{\ketbra22}(R)}{W_0(R)}=1-4R+2R^2.
 \label{eq:s-three-wigner-ratio}
\end{equation}

Before giving the proof, let us explain why a weight of order $\epsilon^4$ on $\ket2$ suffices to restore Wigner positivity.
The two terms in Eq.~\eqref{eq:s-three-wigner-ratio} are negative in two disjoint regions of phase space.
The Wigner function of $\ket{\phi_\epsilon}$ is negative precisely where
\begin{align}
 1-\frac1{\sqrt2}
 <(x+\epsilon^{-1/2})^2+p^2
 <1+\frac1{\sqrt2},
\end{align}
that is, on an annulus centered at $(-\epsilon^{-1/2},0)$ with radii independent of $\epsilon$, and it is bounded below by $-\epsilon^2/(2N_\epsilon)$.
On this annulus, $R=\epsilon^{-1}+O(\epsilon^{-1/2})$ uniformly, so that $\epsilon^4P_2(R)\sim2\epsilon^2$, which is enough to compensate for the negative contribution of $\ket{\phi_\epsilon}$ when $\epsilon$ is small.
Conversely, $P_2(R)$ is negative only on a fixed annulus around the origin, where $\epsilon^4P_2(R)\ge-\epsilon^4$ while $(1-\epsilon^4)W_{\ketbra{\phi_\epsilon}{\phi_\epsilon}}/W_0$ tends uniformly to one.

Let us now make this argument rigorous.
We shall prove that
\begin{equation}
 W_{\sigma_\epsilon}(x,p)>0
 \quad\text{for every }(x,p)\in\R^2
 \quad\text{if }0<\epsilon\le\frac1{100}.
 \label{eq:s-three-explicit-range}
\end{equation}
The value $1/100$ is chosen for convenience and is not optimal.
Multiplying Eq.~\eqref{eq:s-three-wigner-ratio} by the positive factor $N_\epsilon/(1-\epsilon^4)$, it suffices to show that
\begin{equation}
 (q_\epsilon-\epsilon)^2-\frac{\epsilon^2}{2}
 +\frac{N_\epsilon}{1-\epsilon^4}\epsilon^4P_2(R)>0,
 \label{eq:s-three-scaled-polynomial}
\end{equation}
with $1<\frac{N_\epsilon}{1-\epsilon^4}<2$ for $0<\epsilon\le\frac1{100}$.

First consider the fixed annulus where $P_2(R)\le0$.
On this region, $R<2$ and $P_2(R)\ge-1$.
Since $|x|\le\sqrt R$, we have $q_\epsilon=1+2\sqrt\epsilon\,x+\epsilon R\ge(1-\sqrt{\epsilon R})^2$, and hence
\begin{equation}
 \sqrt{q_\epsilon}\ge1-\sqrt{\epsilon R}>
 1-\sqrt{2\epsilon}>\frac45,
 \qquad q_\epsilon-\epsilon>\frac12.
 \label{eq:s-three-q-lower-bound}
\end{equation}
Thus the left-hand side of Eq.~\eqref{eq:s-three-scaled-polynomial} is greater than $1/4-\epsilon^2/2-2\epsilon^4>0$.

Outside this annulus, $P_2(R)>0$, so positivity is immediate unless the other term is negative, that is, $(q_\epsilon-\epsilon)^2-\epsilon^2/2<0$.
In this remaining case,
\begin{equation}
 q_\epsilon<\left(1+\frac1{\sqrt2}\right)\epsilon,
 \qquad
 \sqrt{\epsilon R}\ge1-\sqrt{q_\epsilon}
 >1-\frac43\sqrt\epsilon\ge\frac{13}{15}.
 \label{eq:s-three-s-lower-bound}
\end{equation}
Writing $S=\epsilon R$, we have $S>(13/15)^2>3/4$.
Since $2S^2-4\epsilon S+\epsilon^2$ increases for $S>\epsilon$, it follows that
\begin{equation}
 \epsilon^2P_2(R)=2S^2-4\epsilon S+\epsilon^2
 >2\left(\frac34\right)^2
 -\frac4{100}\frac34>1.
 \label{eq:s-three-scaled-fock-lower-bound}
\end{equation}
The positive term in Eq.~\eqref{eq:s-three-scaled-polynomial} therefore exceeds $\epsilon^2$, while the remaining terms are bounded below by $-\epsilon^2/2$.
This completes the proof of Eq.~\eqref{eq:s-three-explicit-range}.

\subsection{Asymptotic Wigner entropy}
\label{app:s-three-entropy}

Set
\begin{equation}
 \delta_\epsilon=\frac{W_{\sigma_\epsilon}}{W_0}-1,
 \qquad
 \overline n_\epsilon=\Tr(\sigma_\epsilon\hat n).
 \label{eq:s-three-relative-deviation}
\end{equation}
Equation~\eqref{eq:s-entropy-identity} separates the entropy calculation into two terms:
\begin{equation}
 h(W_{\sigma_\epsilon})-h(W_0)
 =2\overline n_\epsilon-D(W_{\sigma_\epsilon}\Vert W_0).
 \label{eq:s-three-entropy-identity}
\end{equation}
The mean photon number follows directly from Eq.~\eqref{eq:s-three-mixture}:
\begin{align}
 2\overline n_\epsilon
 &=(1-\epsilon^4)\frac{4\epsilon+2\epsilon^2}{N_\epsilon}
 +4\epsilon^4 \nonumber\\
 &=4\epsilon-6\epsilon^2+10\epsilon^3+O(\epsilon^4).
 \label{eq:s-three-energy}
\end{align}

To compute the relative entropy up to order $\epsilon^3$, we need the first six powers of $\delta_\epsilon$, since its leading term is of order $\sqrt\epsilon$.
Expanding the ratio in Eq.~\eqref{eq:s-three-wigner-ratio} gives
\begin{align}
 \delta_\epsilon={}&4\epsilon^{1/2}x
 +\epsilon(2R+4x^2-4)
 +4\epsilon^{3/2}x(R-3)\nonumber\\
 &+\epsilon^2(R^2-6R-8x^2+8)
 +2\epsilon^{5/2}x(11-4R)\nonumber\\
 &+\epsilon^3(-2R^2+11R+14x^2-14)
 +O\!\left(\epsilon^{7/2}(1+R^2)\right).
 \label{eq:s-three-deviation-expansion}
\end{align}
The constant in the remainder bound is independent of $x$ and $p$.
We justify the integration of this expansion before evaluating its coefficients.
Because $\delta_\epsilon$ is a polynomial of degree at most four in $x,p$ with coefficients $O(\sqrt\epsilon)$, we have
\begin{equation}
 |\delta_\epsilon(x,p)|\le C\sqrt\epsilon\,(1+R^2)
 \label{eq:s-three-deviation-bound}
\end{equation}
with a constant independent of $x,p$ and sufficiently small positive $\epsilon$.
We expand through degree seven to obtain an integrated remainder $O(\epsilon^4)$, bounded by Eq.~\eqref{eq:s-global-taylor} as
\begin{equation}
 C\int|\delta_\epsilon|^8\dd\nu_1
 \le C'\epsilon^4\int(1+R^2)^8\dd\nu_1
 =O(\epsilon^4).
 \label{eq:s-three-remainder-control}
\end{equation}
Moreover, the ratio is analytic in $\eta=\sqrt\epsilon$ near zero and is invariant under $(\eta,x)\mapsto(-\eta,-x)$.
Gaussian integration therefore removes every odd power of $\eta$ from each moment of $\delta_\epsilon$.
In particular, $\int\delta_\epsilon^7\dd\nu_1=O(\epsilon^4)$ because its only possible term of lower order is proportional to $\epsilon^{7/2}x^7$.
Normalization also gives $\int\delta_\epsilon\dd\nu_1=0$.
It follows that
\begin{equation}
 D(W_{\sigma_\epsilon}\Vert W_0)
 =\sum_{k=2}^{6}\frac{(-1)^k}{k(k-1)}
 \int\delta_\epsilon^k\dd\nu_1+O(\epsilon^4).
 \label{eq:s-three-log-series}
\end{equation}

The Gaussian moments in Eq.~\eqref{eq:s-three-renyi-moments} now determine all coefficients by polynomial multiplication.
For example, the coefficient of $\epsilon^2$ in the second moment contains both square and cross terms:
\begin{equation}
 \int\bigl[(2R+4x^2-4)^2+32x^2(R-3)\bigr]\dd\nu_1
 =20-16=4.
\end{equation}

The full calculation is summarized in Table~\ref{tab:s-three-moments}, and substitution into Eq.~\eqref{eq:s-three-log-series} gives
\begin{equation}
 D(W_{\sigma_\epsilon}\Vert W_0)
 =4\epsilon-6\epsilon^2+\frac{32}{3}\epsilon^3+O(\epsilon^4).
 \label{eq:s-three-relative-entropy-expansion}
\end{equation}

\begin{table}
 \centering
 \renewcommand{\arraystretch}{1.2}
 \begin{tabular}{c@{\hspace{1.2em}}r@{\hspace{1.2em}}r@{\hspace{1.2em}}r}
  \toprule
  $k$ & Coefficient of $\epsilon$ & Coefficient of $\epsilon^2$ & Coefficient of $\epsilon^3$ \\
  \midrule
  $2$ & $8$ & $4$ & $-16$ \\
  $3$ & $0$ & $144$ & $176$ \\
  $4$ & $0$ & $192$ & $4416$ \\
  $5$ & $0$ & $0$ & $11520$ \\
  $6$ & $0$ & $0$ & $7680$ \\
  \bottomrule
 \end{tabular}
 \caption{Coefficients of $\int\delta_\epsilon^k\dd\nu_1$ through order $\epsilon^3$.
 Each row has a remainder $O(\epsilon^4)$.}
 \label{tab:s-three-moments}
\end{table}

The terms of order $\epsilon$ and $\epsilon^2$ cancel those in Eq.~\eqref{eq:s-three-energy}, leaving
\begin{equation}
 h(W_{\sigma_\epsilon})-h(W_0)
 =-\frac23\epsilon^3+O(\epsilon^4).
 \label{eq:s-three-entropy-expansion}
\end{equation}
Thus $\sigma_\epsilon$ is a Wigner-positive counterexample for all sufficiently small positive $\epsilon$.
The range in Eq.~\eqref{eq:s-three-explicit-range} guarantees Wigner positivity, while the entropy expansion establishes the violation for $\epsilon$ small enough.

\section{Wigner--R\'enyi entropies}
\label{app:Renyi}

We now extend both constructions to R\'enyi entropy of every fixed order $0<\alpha<2$, excluding the Shannon case $\alpha=1$ already proven above.
Constants in the remainder estimates may depend on $\alpha$, and also on $n$ for the first family.
We use the Gaussian measure $\dd\nu_\alpha$ and the comparison identity~\eqref{eq:s-renyi-ratio} of Appendix~\ref{app:prelim}.

\subsection{The \texorpdfstring{$0$--$n$}{0--n} family}

Fix $n\ge3$ and let $s(t)$ be either the maximal coherence $s_n(t)$ or the explicit admissible choice $\widetilde s_n(t)$ (see Appendix \ref{app:0n}).
Both satisfy $s(t)^2=t+o(t)$, so the same calculation applies to both states.
Write $\delta_{n,t}=tA_n+s(t)B_n$ as in Eq.~\eqref{eq:s-relative-deviation}.
The Laguerre and angular integrals of Appendix~\ref{app:prelim} give
\begin{align}
 \int_{\R^2}\delta_{n,t}\dd\nu_\alpha
 &=t\left[\left(\frac{2-\alpha}{\alpha}\right)^n-1\right],
 \label{eq:s-renyi-first-moment}\\
 \int_{\R^2}\delta_{n,t}^2\dd\nu_\alpha
 &=t^2\int_{\R^2}A_n^2\dd\nu_\alpha
   +s(t)^2\int_{\R^2}B_n^2\dd\nu_\alpha\nonumber\\
 &=\frac{2^{n+1}}{\alpha^n}t+o(t).
 \label{eq:s-renyi-quadratic-moment}
\end{align}
The mixed term vanishes because $A_n$ is radial and $\int_0^{2\pi}B_n(R,\theta)\dd\theta=0$.

As in Appendix~\ref{sec:s-binary-entropy}, the second-order expansion can be integrated even though $1+\delta_{n,t}$ need not stay away from zero.
Indeed, $\delta_{n,t}\to0$ pointwise, and
\begin{equation}
 \frac{(1+\delta_{n,t})^\alpha-1-\alpha\delta_{n,t}}{t}
 \longrightarrow \frac{\alpha(\alpha-1)}2 B_n^2.
\end{equation}
Equations~\eqref{eq:s-renyi-linear-bound} and~\eqref{eq:s-binary-dominating-bound} bound its absolute value by $C_\alpha(2A_n^2+2B_n^2)$, which is also integrable against $\nu_\alpha$.
Dominated convergence therefore implies
\begin{align}
 \int_{\R^2}(1+\delta_{n,t})^\alpha\dd\nu_\alpha
 =1+\Biggl\{
 \alpha\left[\left(\frac{2-\alpha}{\alpha}\right)^n-1\right]
 +2^n(\alpha-1)\alpha^{1-n}
 \Biggr\}t+o(t).
 \label{eq:s-renyi-integral-expansion}
\end{align}
Taking the logarithm in Eq.~\eqref{eq:s-renyi-ratio} yields
\begin{align}
 h_\alpha(W_{\rho_n(t,s(t))})-h_\alpha(W_0)
 &=C_{n,\alpha}t+o(t),\nonumber\\
 C_{n,\alpha}
 &=\frac{\alpha}{1-\alpha}
 \left[\left(\frac{2-\alpha}{\alpha}\right)^n-1\right]
 -2^n\alpha^{1-n}.
 \label{eq:s-binary-renyi-raw}
\end{align}

To show that this expression is negative, we set $b=(2-\alpha)/\alpha>0$ and compute
\begin{align}
 C_{n,\alpha}
 &=2\frac{b^n-1}{b-1}-2(1+b)^{n-1}\nonumber\\
 &=2\sum_{k=0}^{n-1}b^k
   -2\sum_{k=0}^{n-1}\binom{n-1}{k}b^k\nonumber\\
 &=-2\sum_{k=1}^{n-2}
 \left[\binom{n-1}{k}-1\right]
 \left(\frac{2-\alpha}{\alpha}\right)^k.
 \label{eq:s-binary-renyi-coefficient}
\end{align}
Every term in the final sum is positive for $n\ge3$ and $1\le k\le n-2$.
Thus $C_{n,\alpha}<0$, and both choices of coherence produce a violation for all sufficiently small positive $t$.
At $\alpha=1$, this gives $C_{n,1}=2n-2^n$, matching the derivation of Eq.~\eqref{eq:s-main-expansion}.

\subsection{The \texorpdfstring{$0$--$1$--$2$}{0--1--2} family}
\label{app:renyi-0-1-2}

For $\sigma_\epsilon$, we use the same deviation $\delta_\epsilon=W_{\sigma_\epsilon}/W_0-1$ as in the previous sections.
Only the Gaussian weight and the Taylor coefficients change.
Since $\delta_\epsilon$ starts at order $\sqrt\epsilon$, powers up to six can contribute to the coefficient of $\epsilon^3$.
The remainder and parity estimates of Appendix~\ref{app:s-three-entropy} also apply under the Gaussian weight $\nu_\alpha$, giving an error $O(\epsilon^4)$.
Normalization gives $\int\delta_\epsilon\dd\nu_1=0$, but the first moment need not vanish under $\nu_\alpha$ and must now be retained.
Hence,
\begin{equation}
 \int_{\R^2}(1+\delta_\epsilon)^\alpha\dd\nu_\alpha
 =1+\sum_{k=1}^{6}\binom{\alpha}{k}
 \int_{\R^2}\delta_\epsilon^k\dd\nu_\alpha
 +O(\epsilon^4),
 \label{eq:s-three-renyi-binomial}
\end{equation}
where $\binom{\alpha}{k}=\alpha(\alpha-1)\cdots(\alpha-k+1)/k!$.

Table~\ref{tab:s-three-renyi-moments} gives the required moments, obtained by inserting Eq.~\eqref{eq:s-three-deviation-expansion} and using Eq.~\eqref{eq:s-three-renyi-moments}.

\begin{table}
 \centering
 \renewcommand{\arraystretch}{2.4}
 \small
 \begin{tabular}{c@{\hspace{1.2em}}c@{\hspace{1.2em}}c@{\hspace{1.2em}}c}
  \toprule
  $k$ & Coefficient of $\epsilon$ & Coefficient of $\epsilon^2$ & Coefficient of $\epsilon^3$ \\
  \midrule
  $1$ & $\dfrac{4(1-\alpha)}{\alpha}$
      & $\dfrac{2(1-\alpha)(1-4\alpha)}{\alpha^2}$
      & $-\dfrac{2(1-\alpha)(2-7\alpha)}{\alpha^2}$ \\
  $2$ & $\dfrac8\alpha$
      & $\dfrac{4(4\alpha^2-20\alpha+17)}{\alpha^2}$
      & $\dfrac{16(-4\alpha^3+19\alpha^2-22\alpha+6)}{\alpha^3}$ \\
  $3$ & $0$
      & $\dfrac{48(5-2\alpha)}{\alpha^2}$
      & $\dfrac{16(-4\alpha^3+60\alpha^2-177\alpha+132)}{\alpha^3}$ \\
  $4$ & $0$
      & $\dfrac{192}{\alpha^2}$
      & $\dfrac{192(4\alpha^2-32\alpha+51)}{\alpha^3}$ \\
  $5$ & $0$ & $0$
      & $\dfrac{3840(4-\alpha)}{\alpha^3}$ \\
  $6$ & $0$ & $0$
      & $\dfrac{7680}{\alpha^3}$ \\
  \bottomrule
 \end{tabular}
 \caption{Coefficients of $\int_{\R^2}\delta_\epsilon^k\dd\nu_\alpha$ through order $\epsilon^3$.
 Each row has a remainder $O(\epsilon^4)$.}
 \label{tab:s-three-renyi-moments}
\end{table}

After multiplication by $\binom{\alpha}{k}$ and summation, the coefficients of $\epsilon$ and $\epsilon^2$ cancel exactly.
The remaining coefficient gives
\begin{equation}
 \int_{\R^2}(1+\delta_\epsilon)^\alpha\dd\nu_\alpha
 =1-\frac{2(\alpha-1)(\alpha-2)}{3\alpha}\epsilon^3
 +O(\epsilon^4).
 \label{eq:s-three-renyi-integral-expansion}
\end{equation}
Using $\ln(1+y)=y+O(y^2)$ in Eq.~\eqref{eq:s-renyi-ratio}, we obtain
\begin{equation}
 h_\alpha(W_{\sigma_\epsilon})-h_\alpha(W_0)
 =-\frac{2(2-\alpha)}{3\alpha}\epsilon^3
 +O(\epsilon^4).
 \label{eq:s-three-renyi-expansion}
\end{equation}
The coefficient is negative for every fixed $0<\alpha<2$.
Its value at $\alpha=1$ agrees with the coefficient $-2/3$ obtained in Eq.~\eqref{eq:s-three-entropy-expansion}.
Thus the $0$--$1$--$2$ family also violates the conjectured vacuum bound for all sufficiently small positive $\epsilon$.
For both families, the admissible upper threshold for an entropy violation may depend on $\alpha$.

\end{document}